# Photoinduced Floquet topological magnons in a ferromagnetic checkerboard lattice

Zhiqin Zhang[1,a], Wenhui Feng[1,a], Yingbo Yao[3], Bing Tang[1,2*]

[1] *Department of Physics, Jishou University, Jishou 416000, China*

[2] *The Collaborative Innovation Center of Manganese-Zinc-Vanadium Industrial Technology, Jishou University, Jishou 416000, China*

[3] *College of Information and Electronic Engineering, Hunan City University, Yiyang 413000, China*

ABSTRACT

   This theoretical work is devoted to investigating laser-irradiated Floquet topological magnon insulators on a two-dimensional checkerboard ferromagnet and corresponding topological phase transitions. It is shown that the checkerboard Floquet topological magnon insulator is able to be transformed from a topological magnon insulator into another one possessing various Berry curvatures and Chern numbers by changing the light intensity. Especially, we also show that both Tamm-like and topologically protected Floquet magnon edge states can exist when a nontrivial gap opens. In addition, our results display that the sign of the Floquet thermal Hall conductivity is also tunable via varying the light intensity of the laser field.

* Corresponding author.
E-mail addresses: bingtangphy@jsu.edu.cn
[a] These authors contributed equally to this work.

# 1. Introduction

Up to now, topological insulators have received more and more attention from both theoretical and experimental researchers[1]. Such nontrivial insulators have been realized in a lot of electronic systems, which possess a bulk energy band gap like a common insulator but possess the topologically protected edge (or surface) states because of the bulk-boundary correspondence[2]. Theoretically, the idea of the topological band structure does not depend on the statistical characteristic of the quasiparticle (boson or fermion ) excitations. On the other hand, the notions of Chern number and Berry curvature are able to be given for arbitrary topological energy band construction without respect to the quasi-particle excitations. Naturally, such topological notions can be expanded to those bosonic systems, e. g., photons[3], polarions[4], and magnons[5].

In current years, there has been a growing interest in studies on the topological magnon insulator, which can be viewed as the bosonic counterpart of the topological insulator in electronic systems[6]. Topological magnon insulators have nontrivial magnon energy bands and topologically protected magnon edge states. Initially, Onose *et al.* have detected the magnon Hall effect in the insulating quantum ferromagnet $Lu_2V_2O_7$ with pyrochlore lattice[7]. Their results have showed that a transverse heat current appears in the presence of the longitudinal temperature gradient. Immediately, Matsumoto *et al.* have explained that the emergence of the magnon Hall effect is due to edge magnon currents in the two-dimensional (2D) quantum magnetic systems[8,9]. Later on, Zhang *et al.* [10] have taken cognizance of that this magnon edge current is caused by the nontrivial band topology of quantum ferromagnets, and realized topological magnon insulators for the first time. Physically, topological magnons can appear in the insulating quantum magnets possessing various lattice geometries, however it is very hard to experimentally observe real magnetic materials them. So far, an experimental realization of the topological magnon insulator has been successfully finished by using one quasi-2D kagomé quantum ferromagnetic material Cu(1–3, bdc) [11]. Furthermore, topological magnons have also been investigated in ferromagnets with other lattice structures, such as the honeycomb[12], Lieb[13], star[14], checkerboard[15], and Shastry-Sutherland lattices[16].

In principle, those charge-neutral magnons in the insulating quantum magnet can be regarded as the magnetic dipole moments hopping on lattices[17,18]. A magnetic dipole moving in the external electric field accumulates a geometric phase known as the Aharonov-Casher phase[19], which is analogous to the Aharonov-Bohm phase accumulated via a charge particles in the external magnetic field. This phenomenon referred to as the Aharonov-Casher effect, which provides an effective way to electrically control the magnon propagation[20]. The Aharonov-Casher effect of the magnon has been experimentally observed in a single-crystal yttrium iron garnet[21]. It has been shown that magnons hopping in the time-dependent oscillating electric field can cause some interesting characteristics. Very recently, Owerre has found that hopping magnons in one time-dependent oscillating electric (laser) field can generate the Floquet topological magnon insulator, which is analogous to electronic Floquet topological insulator[22-24]. In his works, the ferromagnets on the kagomé, honeycomb ferromagnet, and Lieb lattices have been considered. He has shown that, in Floquet magnon topological insulators, a chiral or alterable Dzyaloshinskii-Moriya (DM) interaction can be induced via a circular-polarized laser.

In this letter, we will theoretically study laser-irradiated Floquet topological magnon insulators on a ferromagnetic checkerboard lattice and the corresponding topological phase transition by means of the Floquet-Bloch theory. Our idea is to make use of a circular-polarized laser to introduce alterable parameters in checkerboard Floquet topological magnon insulators, which may cause topological phase transitions[24]. By the use of the Floquet-Bloch formalism, we will display that Floquet topological magnon insulators are able to be transformed from a topological magnon insulator into another one, which has various Berry curvatures, and Chern numbers. Especially, edge states in the checkerboard Floquet topological magnon insulator are also to be investigated. Finally, we shall show that the sign of the Floquet magnon thermal Hall conductivity is related to the light intensity of the laser field. More details will be presented in the following sections.

**2. Ferromagnetic checkerboard lattice model**

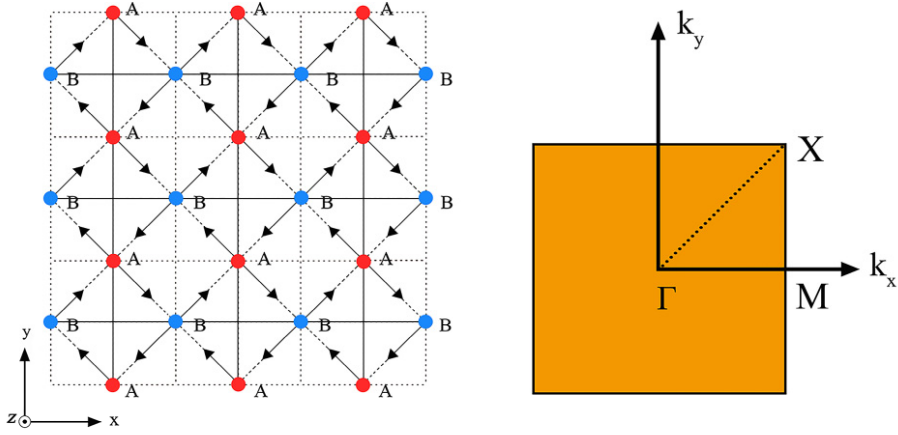

**Fig. 1.** (Color online) The checkerboard lattice (left) and the corresponding first Brillouin zone (right). We have near neighbor interactions $J_1$ and $D$ between local spins at site A and site B, and next near neighbor interaction $J_2$ between lattice site A and A, and B and B, with $J_2$ along the solid line. Sign of $v_{ij}$ is assumed to be positive when the coupling is along the arrow directions and otherwise negative. In the first Brillouin zone, there exists a path marked connecting the $\Gamma$, M, and X points, where $\Gamma = (0,0)$, $M = (\pi,0)$, and $X = (\pi,\pi)$.

In this work, we consider a Heisenberg ferromagnet on a two-dimensional checkerboard lattice. In principle, the checkerboard lattice is able to be viewed as a 2D analog of the 3D pyrochlore lattice[25]. In the context of an applied magnetic field, the Hamiltonian of our checkerboard ferromagnet is given by

$$H = -J_1 \sum_{<i,j>} \vec{S}_i \cdot \vec{S}_j - J_2 \sum_{<<i,j>>} \vec{S}_i \cdot \vec{S}_j + \sum_{<i,j>} \vec{D}_{ij} \cdot (\vec{S}_i \times \vec{S}_j) - g\mu_B \vec{H} \cdot \sum_i \vec{S}_i , \qquad (1)$$

in terms of the local spin operators $\vec{S}_i$. The first two terms in the above equation stand for nearest-neighbor and next-nearest-neighbor ferromagnetic Heisenberg interactions( $J_1 > 0$ and $J_2 > 0$ ), respectively. The third term represents the nearest-neighbor DM interaction, where $\vec{D}_{ij}$ is defined as the DM interaction vector between lattice site $i$ and lattice site $j$. According to Moriya's rules, one can assume $\vec{D}_{ij} = v_{ij}D\vec{e}_z$, where $v_{ij}$ is an orientation-dependent constant [26]. The last term corresponds to a Zeeman coupling with an applied magnetic field $\vec{H} = H_z\vec{e}_z$, where $g$ is called the $g$-factor and $\mu_B = e\hbar/2m_e$ is known as the Bohr magneton. As shown in Fig. 1, there are two inequivalent sites A and B in the ferromagnetic checkerboard lattice. Here, there are the ferromagnetic Heisenberg interaction $J_1$ and the DM interaction $D$ between near neighbor spins at sites A and B. Moreover, the ferromagnetic Heisenberg interaction $J_2$ exist between next near neighbor sites A and A, and B and B.

In order to bosonize the Heisenberg Hamiltonian, we need to employ the Holstein-Primakoff (HP) transformation. Truncated to zeroth order, the HP transformation can be written as $S^z = S - a^+a$, $S^+ \approx \sqrt{2S}a = (S^-)^+$, where $a^+$ ($a$) corresponds to the magnon creation (annihilation) operator, and $S^\pm = S^x \pm iS^y$ stand for the local spin raising and lowering operators. Substituting the zeroth order HP transformation into Eq. (1), one can obtain the following bosonized Hamiltonian

$$H = f_0 \sum_i \left(a_i^+a_i + b_i^+b_i\right) - f_1 \sum_{<i,j>}\left(a_i^+b_j e^{i\varphi_{ij}} + \text{H.c}\right) - f_2 \sum_{<<i,j>>}\left(a_i^+a_j + b_i^+b_j + \text{H.c}\right) \quad (2)$$

with $f_0 = 4J_1S + 2J_2S + g\mu_B H$, $f_1 = S\sqrt{J_1^2 + D^2}$, and $f_2 = J_2S$. The phase $\varphi_{ij} = v_{ij}\varphi = v_{ij}\arctan\left(\dfrac{D}{J_1}\right)$ corresponds to the fictitious magnetic flow in each basic square plaquette of the ferromagnetic checkerboard spin lattice. After performing the Fourier transformation, we can obtain the magnon Hamiltonian in momentum space as $H(\vec{k}) = f_0 I_{2\times 2} - \Lambda(\vec{k})$, where $\Lambda(\vec{k})$ reads

$$\Lambda(\vec{k}) = \begin{pmatrix} 2f_2\cos(k_y) & 2f_1\left[\cos\left(\dfrac{k_x+k_y}{2}\right)e^{-i\varphi} + \cos\left(\dfrac{k_x-k_y}{2}\right)e^{i\varphi}\right] \\ 2f_1\left[\cos\left(\dfrac{k_x+k_y}{2}\right)e^{i\varphi} + \cos\left(\dfrac{k_x-k_y}{2}\right)e^{-i\varphi}\right] & 2f_2\cos(k_x) \end{pmatrix}$$

(3)

with $k_x = \vec{k}\cdot\vec{b}_1$ and $k_y = \vec{k}\cdot\vec{b}_2$. Here, $\vec{b}_1 = (1,0)$ and $\vec{b}_2 = (0,1)$ are the primitive vectors of the checkerboard lattice in Fig. 1. A straightforward diagonalization of the Hamiltonian $H(\vec{k})$ yields the energy spectrum of the single magnon, which contains two bands: the upper and lower bands with the dispersion

$$\varepsilon_{\vec{k}}^{\pm} = f_0 - f_2 t_0 \pm \sqrt{t_1 f_1^2 + t_2^2 f_2^2} \qquad (4)$$

with

$$t_0 = \cos(k_x) + \cos(k_y),$$
$$t_1 = 4\{1 + \cos(k_x)\cos(k_y) + [\cos(k_x) + \cos(k_y)]\cos(2\varphi)\},$$
$$t_2 = \cos(k_x) - \cos(k_y). \qquad (5)$$

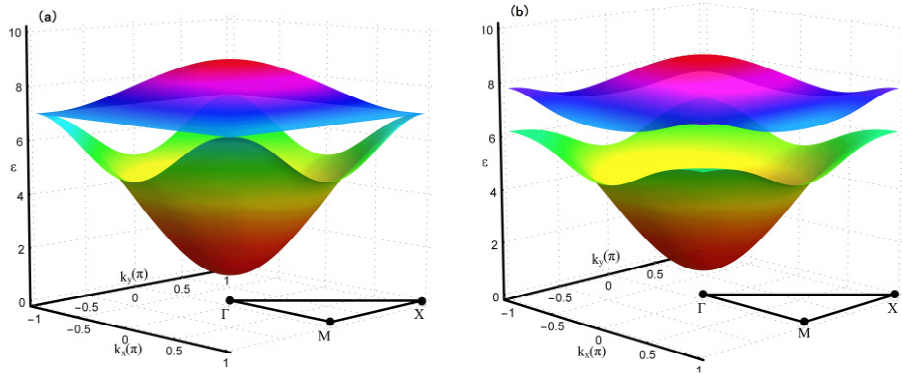

**Fig. 2.** (Color online) The magnon energy bands of the checkerboard ferromagnet: (a) $D = 0$, (b) $D = 0.2$. The other parameters are as follows: $J_1 = 1$, $J_2 = 0.5$, $S = 1$, $H_z = 1$, $g = 1$, and $\mu_B = 1$.

In Fig. 2, we show the complete magnon energy dispersion relation of the present checkerboard ferromagnet. It is obvious that the magnon energy spectrum comprises two bands: the optical "up" and acoustic "down" bands. When the DM interaction is absent, i.e., $D = 0$ or $\varphi = 0$, the optical "up" and acoustic "down" band meet at the X point, as shown in Fig. 2(a). In this case, the checkerboard ferromagnet cannot support the magnon Hall effect. After introducing the DM interaction, the spatial inversion symmetry of the checkerboard spin lattice is broken down, which leads to the emergence of a band gap $\Delta\varepsilon = 8DS$ at the X point as displayed in Fig. 2(b). When the band gap opens, the bulk-boundary correspondence can cause the appearance of the topologically protected magnon (chiral) edge states. Thus, the system becomes a topological magnon insulator. For the checkerboard ferromagnet with the DM interaction, the corresponding Berry curvatures, spin Hall conductivity, spin Nernst coefficients, and magnon thermal Hall conductivity have been calculated in Ref.[15].

**3. Laser-Irradiated checkerboard ferromagnet**

*3.1. The Aharanov Casher phase*

The concept of driven intrinsic topological magnon insulators has been previously introduced in some quantum ferromagnets[22-24]. Physically, magnons are regarded as charge-neutral bosonic quasiparticles. Actually, such bosonic quasiparticles are moving magnetic dipoles in ordered magnetic systems. When the quantization of the spin magnetic dipole moment is parallel to the $z$ direction, the magnetic dipole moment can be represented as $\vec{\mu} = \mu_m \vec{e}_z$ with $\mu_m = g\mu_B$, where $g$ and $\mu_B$ stand for the Landé g-factor and the Bohr magneton, respectively. Here, we assume that magnons in the insulating checkerboard ferromagnetic system are placed in an intense laser field with an oscillating electric field $\vec{E}(\tau)$, which is irradiated perpendicular to the checkerboard ferromagnet lying on the $xy$ plane. According to Owerre's idea, then the oscillating electric field $\vec{E}(\tau)$ can be defined via $\vec{E}(\tau) = -\frac{\partial \vec{A}(\tau)}{\partial \tau}$, where $\vec{A}(\tau)$ corresponds to the time-dependent vector potential of the light (electric) field. In the present work, let us consider a laser field with the electric vector potential $A(\tau) = A_0[\sin(\omega\tau), \sin(\omega\tau + \phi), 0]$, where $A_0$ stands for the amplitude of the electric vector potential, $\omega$ represents the circular frequency of the light, and $\phi$ is the corresponding phase difference. In the present work, we focus on the case of $\phi = \pi/2$, which corresponds to the circularly-polarized light. Obviously, the vector potential

$\vec{A}(\tau)$ possesses temporal periodicity, i.e., $\vec{A}(\tau+T)=\vec{A}(\tau)$, where $T=\frac{2\pi}{\omega}$ is the period. In the presence of this electric field with the temporal periodicity, the magnetic dipole moment of uncharged magnons hopping shall accumulate the Aharanov Casher phase[18,24], namely,

$$\theta_{ij}(\tau)=\frac{g\mu_B}{\hbar c^2}\int_{\vec{r}_i}^{\vec{r}_j}\vec{A}(\tau)\cdot d\vec{l} \quad . \tag{6}$$

Based on Refs. [18, 27], we deduce that the corresponding time-dependent Hamiltonian can be written in the following form

$$H(\tau)=-J_1\sum_{<i,j>}\left[\frac{1}{2}\left(S_i^+S_j^-e^{i\theta_{ij}(\tau)}+S_j^-S_i^+e^{-i\theta_{ij}(\tau)}\right)+S_i^zS_j^z\right]$$
$$-J_2\sum_{<<i,j>>}\left[\frac{1}{2}\left(S_i^+S_j^-e^{i\theta_{ij}(\tau)}+S_j^-S_i^+e^{-i\theta_{ij}(\tau)}\right)+S_i^zS_j^z\right]$$
$$+\frac{iD}{2}\sum_{<i,j>}v_{ij}\left(S_j^+S_i^-e^{i\theta_{ij}(\tau)}-S_i^-S_j^+e^{-i\theta_{ij}(\tau)}\right)-g\mu_B\vec{H}\cdot\sum_i\vec{S}_i.$$

(7)

It is obvious that the time-dependent Hamiltonian in Eq.(7) has the time periodicity $H(\tau+T)=H(\tau)$ with $T=\frac{2\pi}{\omega}$ .

*3.2. Floquet-Bloch formalism*

In order to investigate the periodically driven quantum system in Eq. (7), we shall make use of the Floquet theory [28,29] to change this time-dependent model to a static time-independent effective model, which is governed via the Floquet Hamiltonian. In principle, the static effective Hamiltonian $H_{eff}$ can be written in in terms of $\omega^{-1}$, i.e., $H_{eff}=\sum_{m\geq 0}\omega^{-m}H_{eff}^m$, where $m$ is integer. One can derive the series expansion for the static effective model Hamiltonian by means of the discrete Fourier component of the time-dependent model Hamiltonian $H^n=\frac{1}{T}\int_0^T e^{-in\omega\tau}H(\tau)d\tau$ [24]. Thus, we can obtain

$$H^{(n)} = -J_1 \sum_{<i,j>} \left\{ \delta_{n,0} S_i^z S_j^z + \frac{e^{in\phi_{ij}}}{2} \left[ J_n\left(\frac{\sqrt{2}}{2}A_0\right) S_i^+ S_j^- + J_{-n}\left(\frac{\sqrt{2}}{2}A_0\right) S_i^- S_j^+ \right] \right\}$$
$$- J_2 \sum_{<<i,j>>} \left\{ \delta_{n,0} S_i^z S_j^z + \frac{e^{in\phi_{ij}}}{2} \left[ J_n(A_0) S_i^+ S_j^- + J_{-n}(A_0) S_i^- S_j^+ \right] \right\}$$
$$+ iD \sum_{<i,j>} \frac{v_{ij} e^{in\phi_{ij}}}{2} \left[ J_n\left(\frac{\sqrt{2}}{2}A_0\right) S_i^+ S_j^- - J_{-n}\left(\frac{\sqrt{2}}{2}A_0\right) S_i^- S_j^+ \right] - g\mu_B \vec{H} \cdot \sum_i \vec{S}_i,$$

(8)

where $J_n(x)$ is known as the Bessel function of order $n$ ($n \in \mathbb{Z}$), $\phi_{ij}$ denotes the radian of site $i$ as it rotates counterclockwise along the $x$-axis to site $j$, $\frac{g\mu_B}{\hbar c^2}$ has been absorbed in $A_0$, and $\delta_{n,\ell} = 1$ for $n = \ell$ and zero otherwise. In the above derivation, the mathematical relationship $e^{[izsin(x)]} = \sum_{m=-\infty}^{\infty} J_m(z) e^{imx}$ has been used.

For analytical convenience, let us focus on the magnonic Floquet Hamiltonian corresponding to the off-resonant regime[24]. While the magnon bandwidth $\Delta$ of the undriven system is much less than the driving frequency $\omega$, namely, $\Delta \ll \omega$, this treatment is reasonable. In the off-resonant regime, it suffices to consider the zeroth order of the total effective Floquet Hamiltonian[24]. Thus, we have

$$H_{eff} \approx -\sum_{<i,j>} \left[ J_{1,\perp}\left(S_i^- S_j^+ + S_i^+ S_j^-\right) + J_1 S_i^z S_j^z \right] - \sum_{<<i,j>>} \left[ J_{2,\perp}\left(S_i^- S_j^+ + S_i^+ S_j^-\right) + J_2 S_i^z S_j^z \right]$$
$$+ iD_F \sum_{<i,j>} v_{ij}\left(S_i^+ S_j^- - S_i^- S_j^+\right) - g\mu_B \vec{H} \cdot \sum_i \vec{S}_i,$$

(9)

where $J_{1,\perp} = \frac{1}{2} J_0\left(\frac{\sqrt{2}}{2}A_0\right) J_1$, $J_{2,\perp} = \frac{1}{2} J_0(A_0) J_2$, and $D_F = J_0\left(\frac{\sqrt{2}}{2}A_0\right) \frac{D}{2}$. The corresponding the time-independent Floquet magnon Hamiltonian is given by

$$H_{eff}(\vec{k}) = f_0 I_{2\times 2} - \Lambda'(\vec{k})$$

(10)

with

$$\Lambda'(\vec{k}) = \begin{pmatrix} t_0^{AA} f_2 \cos(k_y) & t_0^{AB} f_1 \left[ \cos\left(\frac{k_x + k_y}{2}\right) e^{-i\varphi} + \cos\left(\frac{k_x - k_y}{2}\right) e^{i\varphi} \right] \\ t_0^{AB} f_1 \left[ \cos\left(\frac{k_x + k_y}{2}\right) e^{i\varphi} + \cos\left(\frac{k_x - k_y}{2}\right) e^{-i\varphi} \right] & t_0^{BB} f_2 \cos(k_x) \end{pmatrix},$$

(11)

where

$$t_0^{AA} = 2J_0(A_0), \quad t_0^{AB} = 2J_0\left(\frac{\sqrt{2}A_0}{2}\right), \quad t_0^{BB} = 2J_0(A_0). \tag{12}$$

Obviously, a direct effect of laser-irradiation is that the magnonic Floquet Hamiltonian (9) is analogous to that of a distorted checkerboard ferromagnet with tunable ferromagnetic Heisenberg and DM interactions. In principle, a Floquet checkerboard topological magnon insulator can be generated due to the existence of the tunable DM interaction. In the following text, let us analyze some topological aspects of the present model.

*3.3. Laser-induced topological transitions*

In the topological systems, the Berry curvature is an very important quantity[2]. Especially, this quantity is the fundamental of lots of observables in those topological insulators. In magnonic systems, a nontrivial band topology can appear only when the system reveals a nontrivial energy gap between two magnon energy bands and one nonzero Chern number suggests the presence of magnon edge states in these systems. In order to investigate the laser-induced topological phase transition in the periodically perturbation magnonic topological insulators, the Berry curvature for a selected magnon band $\beta$ ($\beta = +,-$) can be defined via

$$\Omega_\beta(\vec{k}) = -\sum_{\beta' \neq \beta} \frac{2\text{Im}\left(\langle \psi_{\vec{k},\beta}|\hat{v}_x|\psi_{\vec{k},\beta'}\rangle \langle \psi_{\vec{k},\beta'}|\hat{v}_y|\psi_{\vec{k},\beta}\rangle\right)}{\left(\varepsilon_{\vec{k},\beta} - \varepsilon_{\vec{k},\beta'}\right)^2}, \tag{13}$$

where $\hat{v}_i = \partial H_{eff}(\vec{k})/\partial k_i$ ($i = x, y$) stands for the velocity operator, $\psi_{\vec{k},n}$ represents the magnon eigenstate, and $\varepsilon_{\vec{k},n}$ is the corresponding magnon energy eigenvalue. Furthermore, the linked Chern number can be expressed as a integration with regard to the Berry curvature through the whole first Brillouin zone (BZ), which has the following form

$$C_\beta = \frac{1}{2\pi} \int_{BZ} d^2k \, \Omega_\beta(\vec{k}). \tag{14}$$

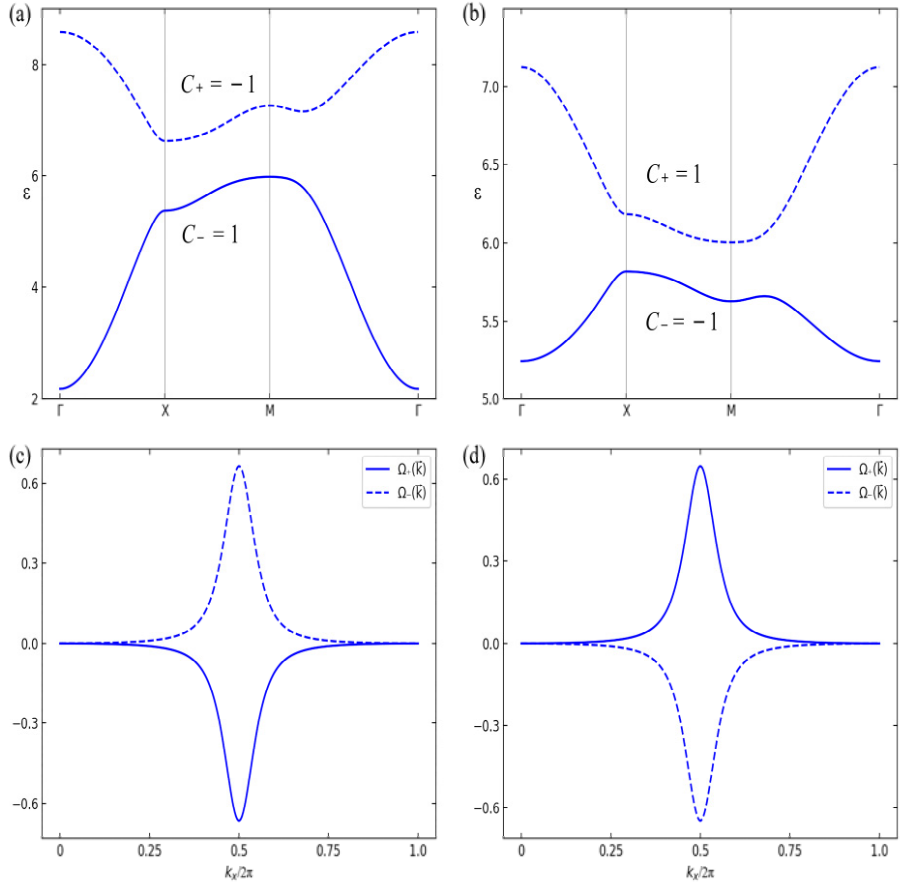

**Fig. 3.** (Color online) Topological magnon bands for the Floquet checkerboard topological magnon insulator (top panel): (a) $A_0 = 1.3$, (b) $A_0 = 2.8$. Tunable Berry curvatures for the Floquet checkerboard topological magnon insulator at $k_y = 0$ (bottom panel): (c) $A_0 = 1.3$, (d) $A_0 = 2.8$. The relevant parameters are set to $J_1 = 1$, $J_2 = 0.5$, $D = 0.2$, $S = 1$, $H_z = 1$, $g = 1$, and $\mu_B = 1$.

In Fig. 3, we display the Floquet topological magnon bands and the Berry curvatures of the Floquet checkerboard topological magnon insulator for different light intensities (i.e., different values of $A_0$). It is clearly seen that the acoustic and optical Floquet topological magnon bands and associated Berry curvatures vary as the light intensity of the laser field changes. Furthermore, Fig. 4 shows the Chern number for the Floquet topological magnon band as a function of the vector potential amplitude $A_0$. It is obvious that the topological magnonic system can be switched from one checkerboard Floquet topological magnon insulator possessing Chern numbers $(C_+, C_-) = (-1, 1)$ to another one possessing Chern numbers $(C_+, C_-) = (1, -1)$. In fact, such results manifest that changing the light intensity of the laser field can redistribute the magnon energy band constructions of a checkerboard Floquet topological magnon insulator and subsequently causes the topological phase transition from one checkerboard Floquet topological magnon insulator to another one possessing various Berry curvatures and Chern numbers.

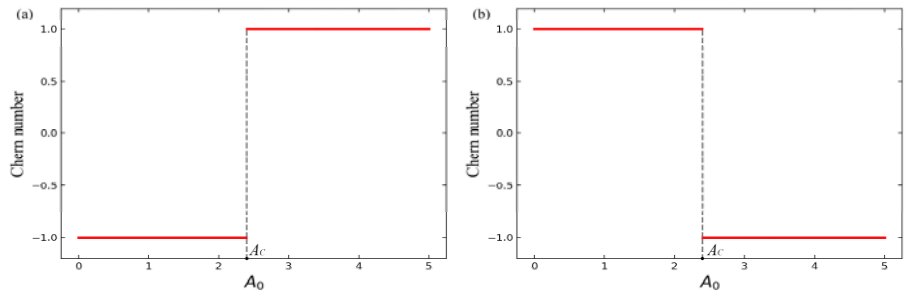

**Fig. 4.** The Chern number for the checkerboard Floquet topological magnon band versus the electric vector potential amplitude $A_0$: (a) the optical "up" band, (b) the acoustic "down" band. Here, the critical value $A_c$ is approximately equal to 2.40. The other parameters are the same as in Fig. 3.

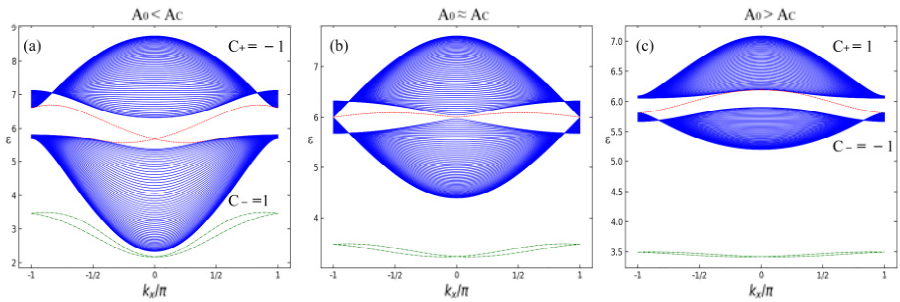

**Fig. 5.** Floquet magnon energy dispersions of edge and bulk states in the laser-irradiated checkerboard ferromagnet with a zig-zag edge at several values of $A_0$. The other parameters are the same as in Fig. 3.

Physically, magnon edge states can exist due to the nontrivial topology of the Berry curvatures. Here, let us consider a zigzag boundary. By use of the numerical diagonalization technique, we can calculate the Floquet magnon energy spectrums of a zigzag checkerboard ferromagnet at different light intensities, as shown in Fig. 5.

Under the circularly polarized laser possessing high enough frequency, the $n=0$ order of the total effective Floquet Hamiltonian yields $H_{DM}^{eff} = D_F \sum_{<i,j>} \vec{e}_z \cdot (\vec{S}_i \times \vec{S}_j)$. Evidently, the intensity of the intrinsic DM interaction can be manipulated via changing the value of $A_0$. In the case of $A_0 < A_c \approx 2.40$ an energy gap appears the $X$ point and it closes in the vicinity of the first null point of the Bessel function $A_0 \approx A_c$ and reappears for the case of $A_0 > A_c$. In Fig. 5, we can see clearly that the topologically protected chiral magnon edge state can appear within the bulk energy band gap in the case of $A_0 < A_c$ or $A_0 > A_c$. Furthermore, we note that there always exists Tamm-like edge states[30] below the lower band. However, they correspond to topologically trivial edge modes, which are not of interest in the present work.

*3.4. Laser-induced magnon thermal Hall effects*

A significant manifestation of Floquet topological magnon insulators is identified as the Floquet magnon thermal Hall effect. As a matter of fact, the Floquet magnon edge states causes a transverse heat current when one longitudinal temperature gradient exists. Such Floquet magnon edge flux can give rise to the Floquet magnon thermal Hall effect. The thermal Hall conductivity is an important physical quantity to describe the Floquet magnon thermal Hall effect. Analogous to the electronic Hall conductivity, the magnon thermal Hall conductivity is in connection with the Berry curvature of the Floquet magnon eigenstates[24]. If we only consider the limitation where the Bose distribution function tends to equilibrium, in those nonequilibrium magnonic Floquet system, then the same physical conception of the undriven magnon thermal Hall effect is able to be extended to the laser-induced Floquet topological magnonic system. Thus, the transverse component $\kappa_{xy}$ of the Floquet magnon thermal Hall conductivity can be written in the following form[22]

$$\kappa_{xy}^F = -k_B^2 T \int_{BZ} \frac{dk^2}{(2\pi)^2} \sum_{\beta=\pm} c_2(n_\beta) \Omega_\beta(\vec{k}) \qquad (15)$$

where $n_\beta = \left(e^{\varepsilon_\beta(\vec{k})/k_B T} - 1\right)^{-1}$ correspond to the Bose distribution function tending to thermal equilibrium state, $k_B$ represents Boltzmann's constant, $T$ denotes the absolute temperature, and $c_2(x) = (1+x)\left(\ln\frac{1+x}{x}\right)^2 - (\ln x)^2 - 2\text{Li}_2(-x)$ with $\text{Li}_2(x)$ denoting a double logarithm. Obviously, the Floquet magnon thermal Hall conductivity is considered to be the Berry curvature weighed via the double logarithmic function $c_2$. As a consequence, any variation in the Berry curvature can influence the Floquet magnon thermal Hall conductivity. Fig. 6 shows that the two laser-induced phases in the Floquet topological magnon insulator possess opposite signs of the Floquet magnon thermal Hall conductivity since the sign of the associated Berry curvatures varies. Furthermore, we note that the sign of $\kappa_{xy}^F$ is in accordance with the sign of the Berry curvature or Chern number of the optical "up" Floquet magnon band at low temperatures.

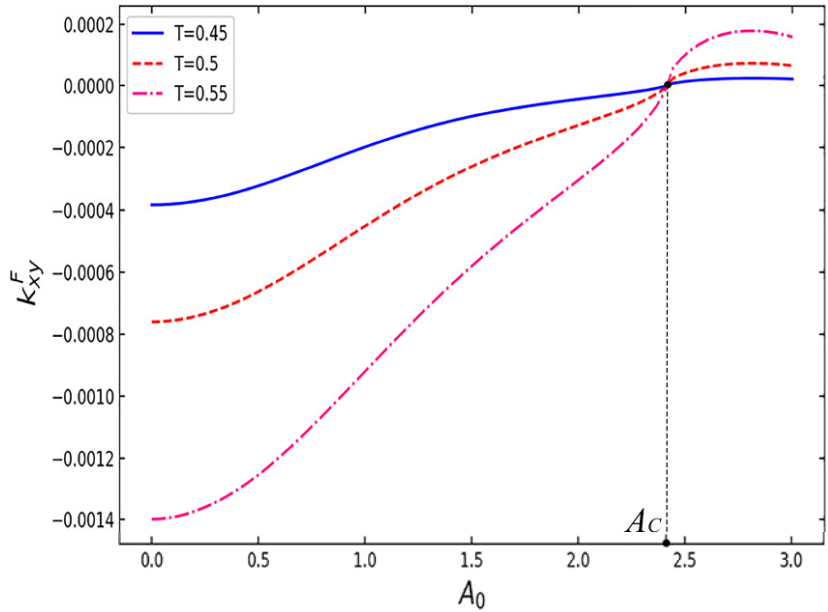

**Fig. 6.** The Floquet thermal Hall conductivity $\kappa_{xy}^F$ versus the electric vector potential amplitude $A_0$ for different temperature values. Here, the Boltzmann constant is fixed as $k_B = 1$, and the other parameters are the same as in Fig. 3.

## 4. Conclusions

In summary, we have theoretically a theoretical study on laser-irradiated Floquet topological magnon insulators on a ferromagnetic checkerboard lattice and the corresponding topological phase transition with the help of the Floquet-Bloch theory. It was shown that a checkerboard Floquet topological magnon insulator can be generated on account of the occurrence of the tunable DM interaction induced by a circular-polarized laser. Our results displayed that the checkerboard Floquet topological magnon insulator is able to be transformed from a topological magnon insulator into another one possessing various Berry curvatures and Chern numbers. Physically, magnon edge states can appear owing to the nontrivial topology of Berry curvatures. We found that both topologically protected and Tamm-like edge states can appear in the magnon band structure. In addition, we showed that the sign of the Floquet magnon thermal Hall conductivity is in accordance with the sign of the Berry curvature or Chern number of the "up" Floquet magnon band at low temperatures.

**Acknowledgments**

This work was supported by the National Natural Science Foundation of China under Grant No. 12064011, the Scientific Research Foundation of Hunan Provincial Education Department under Grant No. 18C0844, the Natural Science Fund Project of Hunan Province under Grant No. 2020JJ4498 and the Graduate Research Innovation Foundation of Jishou University under Grant No. JGY202029.